\documentclass[sigconf,nonacm]{acmart}

\usepackage{tabularx} % Required for auto-wrapping columns
\usepackage{ragged2e} % Recommended for better centering in X columns
\usepackage{booktabs} % Optional: makes lines look professional
\usepackage[table]{xcolor} % Required for row colors

\newcolumntype{M}{>{\centering\arraybackslash}X}

\usepackage{url}
\usepackage{xurl}
\usepackage{graphicx}
\usepackage{textcomp}
\usepackage{listings}

\usepackage{pmboxdraw}

\AtBeginDocument{%
  }

\setcopyright{none}
\newcommand{\name}{ARTHUR}

\begin{document}

%%
%% The "title" command has an optional parameter,
%% allowing the author to define a "short title" to be used in page headers.
\title{Is there a future for models in the LLM era?}

%%
%% The "author" command and its associated commands are used to define
%% the authors and their affiliations.
%% Of note is the shared affiliation of the first two authors, and the
%% "authornote" and "authornotemark" commands
%% used to denote shared contribution to the research.

\author{Marco Calamo}
\email{calamo@diag.uniroma1.it}
\orcid{0009-0006-2602-9604}
\affiliation{%
  \institution{Dipartimento di Ingegneria Informatica Automatica e Gestionale Antonio Ruberti \\ Sapienza Università di Roma}
  \city{Rome}
  \country{Italy}
}

\author{Massimo Mecella}
\email{mecella@diag.uniroma1.it}
\orcid{0000-0002-9730-8882}
\affiliation{%
  \institution{Dipartimento di Ingegneria Informatica Automatica e Gestionale Antonio Ruberti \\ Sapienza Università di Roma}
  \city{Rome}
  \country{Italy}
}

\author{Monique Snoeck}
\email{monique.snoeck@kuleuven.be}
\orcid{0000-0002-3824-3214}
\affiliation{%
  \institution{Research Centre for Information Systems Engineering (LIRIS)\\ KU Leuven}
  \city{Leuven}
  \country{Belgium}
}

%%
%% By default, the full list of authors will be used in the page
%% headers. Often, this list is too long, and will overlap
%% other information printed in the page headers. This command allows
%% the author to define a more concise list
%% of authors' names for this purpose.
\renewcommand{\shortauthors}{Calamo, Mecella, and Snoeck}

%%
%% The abstract is a short summary of the work to be presented in the
%% article.
\begin{abstract}
In an era where software development is deeply tied with Large Language Models, does Model Driven Engineering (MDE) still make sense? This raises the question of the extent to which MDE can be successfully combined with an LLM approach to address the downsides of each approach separately. In this paper, we try to answer that question by investigating a central Research Question: Can Agents powered by Large Language Models improve code generated from UML diagrams and text specifications by Model Driven Engineering (MDE) tools? To investigate this problem, we developed a novel LLM-powered Multi-Agentic approach called ARTHUR (Architecture Refactoring Through Hybrid UML Reasoning), a framework that aims at combining the reliability of MDE and the ease of use of LLMs. ARTHUR is designed to refactor legacy Java code produced by traditional, rule-based, MDE code generators from UML diagrams. To assess this question, we refactored the legacy code of several projects from our custom dataset crafted for this purpose. \name{} made it possible to add support for modern frameworks like Spring Boot, while ensuring compliance with Model-Based Testing techniques to verify that the code still corresponds to the initial model's specifications. We then measured the results obtained in terms of time and cost, passing test rate, and \texttt{compile@k}, \texttt{pass@k} and \texttt{pass$^k$} metrics. We also observed the effect of generating code directly from the conceptual model without the refactoring. Our preliminary test results show that MDE could not be further from retirement, after all.
\end{abstract}

%%
%% Keywords. The author(s) should pick words that accurately describe
%% the work being presented. Separate the keywords with commas.
%% A "teaser" image appears between the author and affiliation
%% information and the body of the document, and typically spans the
%% page.
\begin{teaserfigure}
  \includegraphics[width=\textwidth]{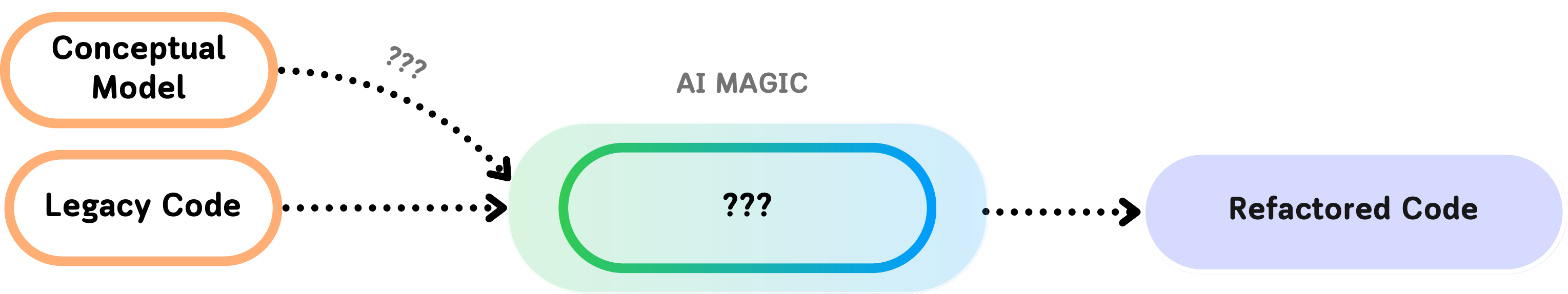}
  \caption{Can we combine the reliability of MDE with usability of LLMs to refactor Java legacy code?}
  \Description{Exploring what AI can do in code refactoring}
  \label{fig:teaser}
\end{teaserfigure}

%%
%% This command processes the author and affiliation and title
%% information and builds the first part of the formatted document.
\maketitle

\section{Introduction}
Model Driven Engineering (MDE) focuses on using models to drive the development of software applications. After its first formalization in 2002~\cite{kent2002model}, the approach gained traction and began to be adopted in various real-world industrial contexts, reaching its peak adoption in the early 2010s~\cite{hutchinson2011model}. However, in the years leading up to the 2020s, MDE gradually lost popularity among practitioners, despite remaining a very successful approach when adopted correctly ~\cite{verbruggen2023practitioners}. The reasons for this inversion of trend are numerous and difficult to pinpoint. Surely, one big factor is the shift away from strictly object-oriented programming in favor of script-like programming, as promoted by languages such as JavaScript or Python~\cite{johri2018identifying}. Nevertheless, we believe that the reason that MDE is no longer the industry standard is its intrinsic complexity. Crafting a \emph{good} conceptual model from a client's requirements, is a very hard task, and fewer and fewer professionals are capable of doing so.

On the other hand, since the mid 2020s, we witness the emergence in software development of the adoption of Large Language Models (LLMs), especially in the new agentic modalities~\cite{watanabe2025use,jiang2026survey}. Their usage makes writing code more efficient and faster. Still, while LLM's problem solving capabilities have proven impressive for single tasks~\cite{chen2021evaluating}, they are still lackluster for full, complex applications~\cite{jin2024llms}. 

\begin{center}
    \begin{figure}[ht]
        \centering
        \includegraphics[width=\linewidth]{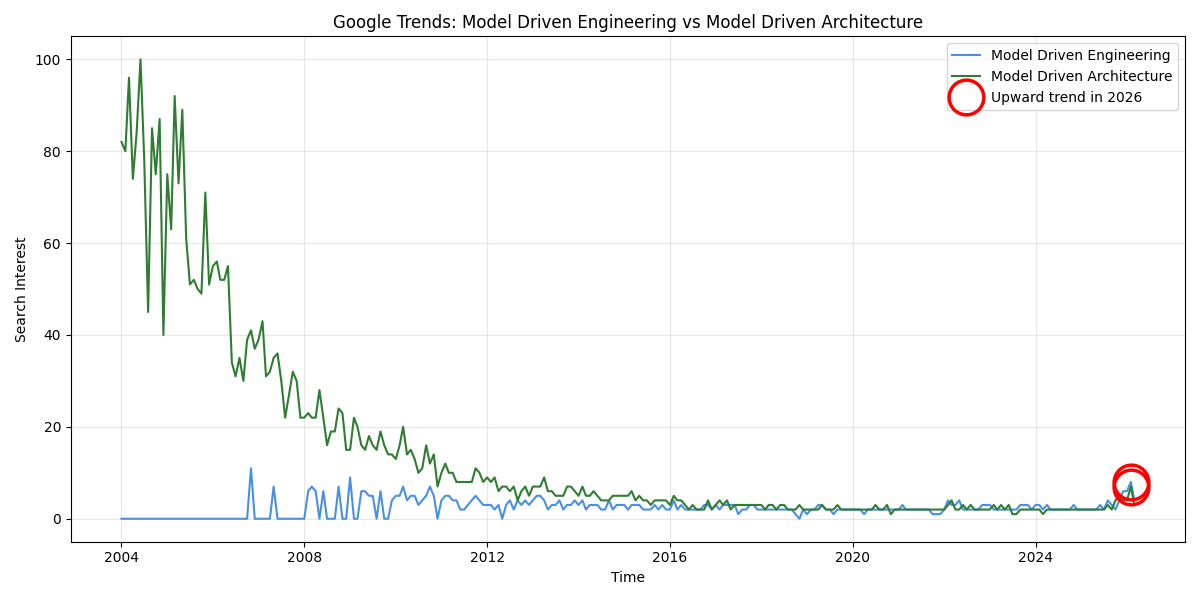}
        \caption{Google Trends on MDE and MDA}
        \label{fig:trend}
        \Description{Trends start to go up again for the first time}
    \end{figure}
\end{center}

From a Google Trends research made at the beginning of March 2026 and reported in Figure~\ref{fig:trend}, it seems that for the first time in a decade, the interest for Model Driven Engineering and Model Driven Architecture (MDA) is increasing, probably as a possible solution on how to guide LLMs in generating complex software. 

This raises the question of the extent to which MDE can be successfully combined with LLM-based approaches to software development; the goal of this paper is to answer the following general research question:

\begin{center}
    \textbf{RQ: Can agents powered by Large Language Models improve code generated from UML diagrams and text specifications by Model Driven Engineering (MDE) approaches and tools?}
\end{center}

To investigate this problem, we developed a novel LLM-powered Multi-Agentic approach called \name{} (Architecture Refactoring Through Hybrid
UML Reasoning), that aims at combining the reliability of MDE and the ease of use of LLMs.  %receiving as input a system's textual requirements and the corresponding UML diagrams, aims at leveraging existing tools that generate Java code from UML to create modern applications ready to be deployed with new technology stacks (such as Spring Boot and React) and different programming languages (such as Python).
\name{} receives the system's textual requirements, the corresponding UML diagrams, and the Java legacy  code generated using existing MDE tools as input and aims to leverage LLMs to create modern applications ready to be deployed with new technology stacks, such as Spring Boot and React. The general idea is to investigate to what extent agents maintains the structure and the constraints of the MDE-generated code. Finally, to ensure the rigor of the testing approach, we want to base it on a the Model-Driven Testing (MDT) approach proposed in~\cite{mussa2009survey}. This allows to be sure that the generated code is still as compliant to the conceptual as the one crafted by the classic existing MDE tools.
%\name{} not only promises not altering the structure and the constraints of the code generated with existing MDE tools, but also aims at generating the missing business logic code that classic code generator do not take care of. Finally, we want to test the generated code using the methodology proposed in Model-Driven Testing~\cite{mussa2009survey}, to be sure that the generated code is still as compliant to the conceptual as the one crafted by the classic existing MDE tools.

%name command is automatic

The following sections of this paper are as follows: Section \ref{sec:rw} presents the work in the literature related to code generation with Model Driven Development methodologies in mind; Section \ref{sec:mt} will present \name{} framework and the experimental setup; Section \ref{sec:rs} will present the results of our tests; finally, Section~\ref{sec:ds} and and Section~\ref{sec:fw} will deeply discuss the experimental results and conclude the paper.

\section{Backgroound and Related Works} \label{sec:rw}
\subsection{Background}

For the Model-Driven Engineering (MDE) and Model-Driven Development (MDD) approaches, the standard modeling language of choice was mainly Unified Modeling Language (UML)~\cite{kent2002model}, for its incredible versatility and support to different kinds of diagrams~\cite{fowler2018uml}. UML has historically been the norm, but modern practices often privilege UML textual representations~\cite{romeo2025uml} like PlantUML\footnote{\url{https://plantuml.com/guide}} or its variant, i.e. tUML \cite{buchmann2024white}. Translating these conceptual architectures into code, to kickstart developers' tasks, relies on generation toolchains such as the Eclipse Modeling Framework (EMF) \footnote{\url{https://eclipse.dev/emf/}} or the Merlin Prototyper\footnote{\url{https://merode.econ.kuleuven.be/pages/CodeGeneration.html}}, which transforms UML diagrams into functional Java applications. 

Recently, a paradigm shift has occurred with the integration of Large Language Models (LLMs) and new agentic modalities. Models such as Claude 4.6, GPT-5, and Kimi 2.5 provide native agentic support capable of complex programming tasks \cite{plaat2025agentic}. To rigorously evaluate code generated by these probabilistic models, researchers utilize metrics like \texttt{pass@k}, which measures the frequency of code passing functional tests within $k$ attempts, alongside \texttt{compile@k} to verify syntactical correctness \cite{yu2025measuring}.

\subsection{Related Works}
To find existing research on solutions that combine LLMs and MDE to generate or alter code, we conducted the following query in Scopus and Google Scholar at the beginning of February 2026:
\begin{center}
    ("LLM" OR 'LARGE LANGUAGE MODEL') AND ('UML' OR 'Model-Driven Development' OR 'REQUIREMENTS ENGINEERING') AND ('CODE GENERATION')
\end{center}
Raw results are available at: \url{https://zenodo.org/records/19222552}.
We then manually filtered and classified the resulting articles by focusing on several categories to identify gaps that can be addressed by the development of \name{}. For each paper we characterized the work done according to the following dimensions. First, we consider the target of work: were LLMs used to generate code to solve a single task (e.g. to make function execution more efficient) or was a full standalone application targeted? Next, we examine whether they used UML diagrams and to what extent, the programming language of the generated the code, and which LLM tool was used. We also considered if the research targeted a specific technology stack in the generation process (e.g. Spring Boot or React). Finally, we extracted how the generated code was evaluated, and whether the code base is publicly available. Results are summarized in Table~\ref{tab1}.

\begin{table*}[htbp]
\footnotesize
\caption{Related Work}
\label{tab1}
\centering

% Start alternating colors on row 2 (the first data row)
% Odd rows will be light gray (gray!10), even rows will be white
%\rowcolors{2}{gray!10}{white}

\begin{tabularx}{\textwidth}{l M M M M M M}
\toprule 

% Force the header row to be white so it doesn't accidentally get colored
\rowcolor{white} 
\textbf{Ref.} & \textbf{Target} & \textbf{UML} & \textbf{Code} & \textbf{Patterns/ Tech} & \textbf{Evaluation} & \textbf{Open Source}\\
\midrule 

\cite{wan2025structure} & Single Task (HumanEval) & AI Generated, activity diagram & GPT-4, Gemma2, Mistral, CodeLLama generated Python C++ Java JavaScript code & -- & pass@1 & \url{https://github.com/SEDevSys/StructGen}\\

\cite{han2024archcode} & Single Task (HumanEval) & -- & GPT-4 generated python code & -- & pass@k (1,2,5) & \url{https://github.com/ldilab/ArchCode} \\

%\cite{north2024code} & Single Task (HumanEval) & -- & WizardCoder, CodeLLama generated Python code & -- & pass@k & --\\

\cite{escobar2025prompt} & Application & Human Generated, class diagram & GPT-4o generated Java code with UML in context & -- & Structural accuracy, compilation, time & Prompts available \\

\cite{buchmann2025model} & Application & Custom UML (tUML) & Local AI (Ollama) generated Java & -- & -- & \url{https://github.com/tbuchmann/mdellmprocessor}\\

\cite{sadik2024llm} & Application & Human Generated, class/activity & GPT-4 generated Java Python code & Java Jade / Python Pade & Cyclomatic complexity & --\\

%\cite{wei2024requirements} & Application & -- & GPT-4 generated Java code & -- & -- & -- (link inactive)\\

\cite{zhao2024generating} & Application & GPT-3.5 class/activity & GPT-3.5 generated Java code & Spring Boot & -- & --\\

\cite{cheon2025llms} & Application & Human generated (UML/OCL) & GPT-4o generated JS code & Dart / Flutter & Code size and complexity, expert evaluation & -- \\

\cite{santos2025applying} & Application & Human generated Class diagram & Human + ChatGPT Javascript code & React & ISO/IEC 25010 metrics & -- \\

\cite{he2025hybrid} & Application & LLM generated class diagram & LLM generated code & -- & pass@k, compile@k (1,3) & \url{https://github.com/zzzzzzaplle/iEcoreGen}  \\

\bottomrule 
\end{tabularx}
\end{table*}

In two out of the nine selected articles, LLMs are leveraged to improve the code MDE-generated from a simple textual description to perform a single task. In this case the improvement should be merely intended in terms of number of test cases passed, since this is the main metric evaluated in the proposed benchmark datasets. Authors of \cite{wan2025structure} demonstrated the impact of adding a LLM-powered agentic pipeline with two actors: a Designer agent to craft a UML activity diagram corresponding to the textual specification to improve the results of the Coder agent on the \texttt{pass@1} metric on several benchmark datasets like HumanEval~\cite{chen2021evaluating}. The \texttt{pass@k} metric checks how often code
passes tests within k attempts, and is one of the most used metrics used for evaluating the quality of LLM-generated code~\cite{yu2025measuring}. The authors of~\cite{han2024archcode} showed that adding requirements, even if generated by LLMs, to the problem description given as input to the LLMs for generating single task code improves the quality of metrics like \texttt{pass@1}. This makes a single generation of the code more reliable.

The majority of works that include conceptual modeling in their code generation pipeline target complex applications and show that adding a rigorous structure to LLMs yields the most benefits. In~\cite{escobar2025prompt}, the authors show that LLMs models starting from the end of 2024, like Gemini 1.5~\footnote{\url{https://storage.googleapis.com/deepmind-media/gemini/gemini_v1_5_report.pdf}} or GPT-4o~\cite{hurst2024gpt} are capable of almost flawlessly translating UML class diagrams into simple yet functional code that represents almost 100\% of the requirements specified in the class diagram (still not achieving a fully functional application). In~\cite{buchmann2025model}, the authors present MoProCo, a Visual Studio Code extension~\footnote{\url{https://marketplace.visualstudio.com/vscode}} that enables LLMs to support both creating a conceptual model in tUML~\cite{buchmann2024white} (a more verbose variant of PlantUML) to represent conceptual models as text, and generating the corresponding Java code. They claimed in the article that MoProCo should seamlessly and intuitively assist the software developer with the aid of LLMs, wether they wanted to produce the tUML model, to code the application or to simply prompt the LLMs for more specific requests, while still being at the center of the generation loop. However, no rigorous validation is provided with this tool. In~\cite{sadik2024llm}, the authors use the test case of a Unmanned Vehicle Fleet for evaluating the capabilities of GPT-4 in generating both Java and Python code in specific Jade~\footnote{\url{https://jade-project.gitlab.io/}} and Pade~\footnote{\url{https://pade.readthedocs.io/}}frameworks using an MDD approach. The authors add UML class and activity diagrams crafted by experts in the generation context to measure the quality of the outputted code. However they only measured the cyclomatic complexity of the code and manually checked its integrity without aspecific metrics. The XdCodeGen approach for generating Java code with LLMs is presented in ~\cite{zhao2024generating}. It consists of four phases: requirement analysis, modeling, code generation, and code verification. The first three of these are supported by GPT-3.5, while the testing uses MSVL~\cite{wang2017msvl}, a custom language designed for testing software. They were able to successfully generate a Java Spring Boot app totaling around 20,000 lines of code that passed the testing stage. However the test was conducted on this task only. The author of~\cite{cheon2025llms} was successfully able to generate a mobile application with the Dart~\footnote{\url{https://dart.dev/}} and Flutter~\footnote{\url{https://flutter.dev/}} frameworks, using GPT-4o with plant UML in context. The evaluation was conducted on code size and complexity in comparison to an equivalent manually written application. The work in~\cite{santos2025applying} investigated how LLMs impact the various stages of software development, including requirements elicitation, architecture design, diagram creation, and implementation. They used a test case application of an online fashion retailer called LUNA and showed that adding LLMs could significantly speed up the development of an application from scratch at the cost of exhibiting higher complexity in certain situations, increased code smells, and lower documentation coverage. 

The work closest to our research question is~\cite{he2025hybrid}. This work proposed the iEcoreGen framework that integrates the Eclipse Modeling Framework (EMF) and LLMs for generating complex software applications. They ran several experiments to add business support to the generated legacy code, tested with a custom dataset. Despite having conducted a rigorous analysis of several approaches and different Language models, the tests were not guided by a Model Driven Testing approach, and there is no capability of supporting different technology stacks or programming languages. In addition most approaches are not open source.

\paragraph{Research Gap}
As can be seen from the overview, most of the approaches target a single case, lack rigorous testing, lack a customizable technology stack, and do not measure the impact of adding models. Based on this, the general research question is refined as follows: 
\begin{itemize}
    \item \textbf{Updated} RQ1: Is it possible to use Agents powered by Large Language Models to improve code generated from UML diagrams and text specifications by existing tools like Merlin, \emph{by supporting different programming languages and technology stacks}?
    \item \emph{RQ 2: How will the generated code behave when tested using Model-Based Testing and modern metrics like \texttt{compile@k} and \texttt{pass@k}/$pass^k$?}
    \item RQ3: What is the added value of incorporating a model as input for the generation process?
\end{itemize}

\section{Methodology}\label{sec:mt}
\begin{center}
    \begin{figure*}[ht]
        \centering
        \includegraphics[width=0.8\linewidth]{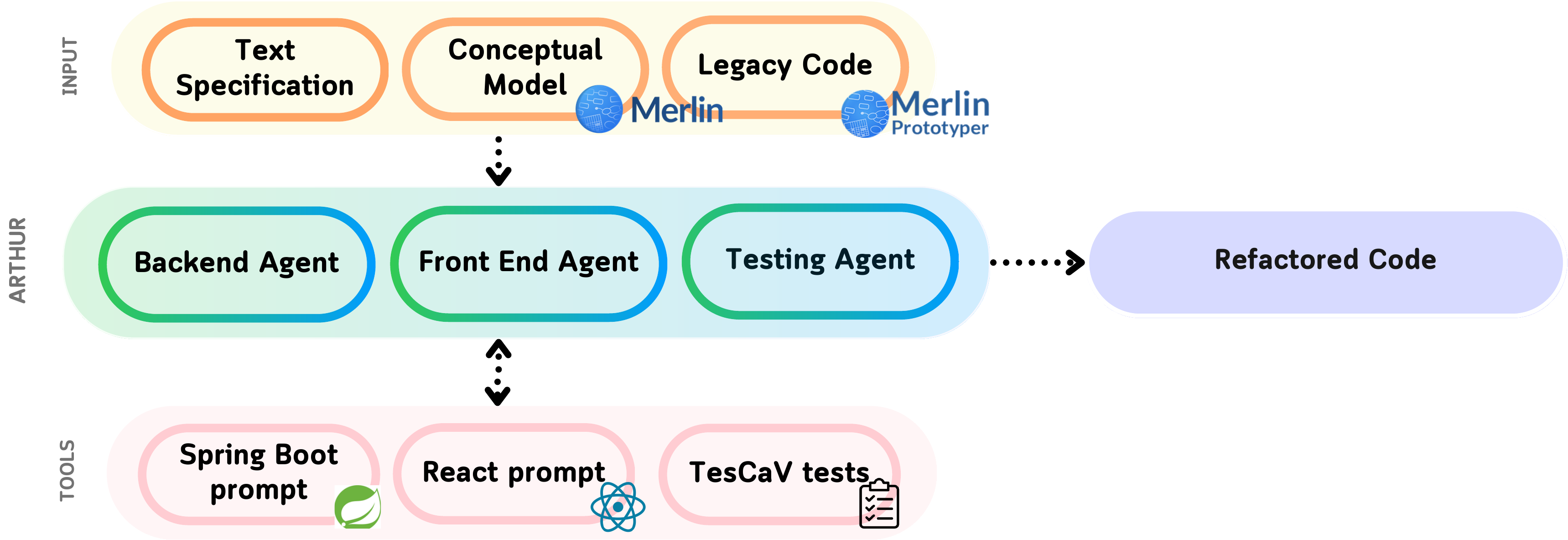}
        \caption{\name{} Architecture}
        \label{fig:archi}
        \Description{The architecture layout}
    \end{figure*}
\end{center}

Since there is no tool in the literature that could satisfy all of our requirements, we needed to develop one from scratch: \name{}. The goal of this paper is not to validate \name{}, but rather run empirical test to seek the answers to our research questions.
The full \name{} architecture is presented in Figure~\ref{fig:archi}. We will proceed to explain the base workflow, breaking it down in its core components, and present the experiment's methodology. The anonymized repository, with a fully functional demo, is available at \url{https://anonymous.4open.science/r/ARTHUR-F2B1/}. The full anonymized generated artifacts are available anonymously on Zenodo at \url{https://zenodo.org/records/19222552}.

\subsection{Dataset and MDE Code Generator}
Before presenting our Multi Agentic system, given the focus on MDE, we need to define exactly what kind of input we need to obtain the best results. To better comprehend how the conceptual models and legacy code are shaped in our dataset, and in what the refactoring consists, we will present a brief overview of how every dataset entry was crafted.

\begin{figure}[ht]
    \centering
    \includegraphics[width=\linewidth]{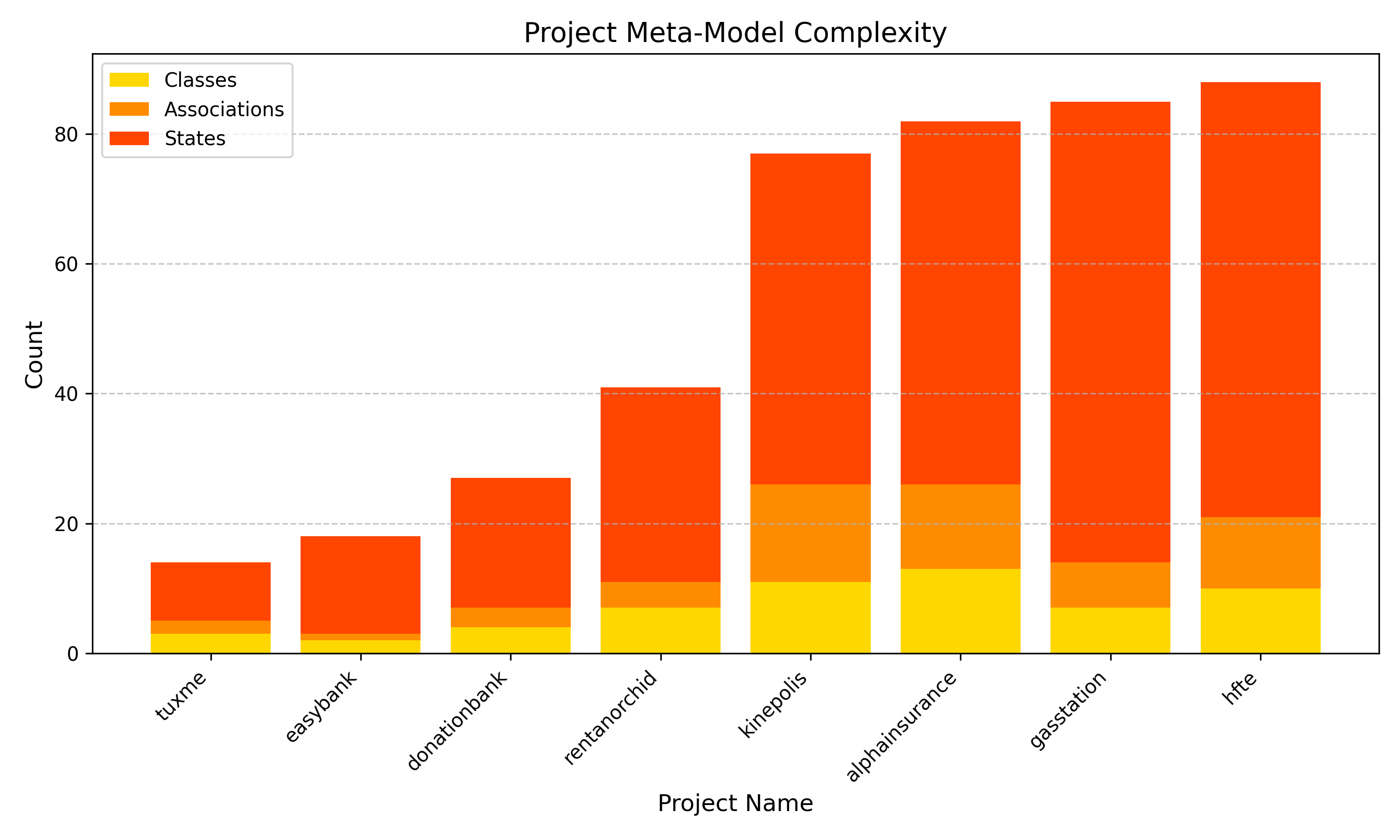}
    \caption{Dataset Complexity Overview in Terms of Number of Classes, Associations and States}
    \label{fig:dataset}
    \Description{Dataset complexity overview}
\end{figure}

First, we needed to identify some business cases with corresponding conceptual models available. To build our test dataset, we carefully selected eight cases of increasing complexity from the Golden UML dataset~\cite{verbruggen2025toward}. Our selection criteria were increasing complexity in terms of number of classes, associations and availability of a conceptual model compatible with our designed code generator. The final eight cases that we selected are broken down by complexity in Figure~\ref{fig:dataset}. 

Every project in the dataset is characterized by a business case description written in natural text, where all requirement and logic are clearly stated. The description is paired with a human-made conceptual model expressed in PlantUML.  The quality of these models and their exact correspondence to the description is ensured by the fact that the selected cases were used for more than 10 years with groups of more than 100 students yearly. During these years, the case descriptions underwent refinement and polishing until the case text was sufficiently clear and complete to guarantee an unambiguous specification of the desired model solution. Before uploading to the Golden UML dataset, non-author of the cases further verified that the models were correct, complete, and consistent with respect to the descriptions. 

To make the models usable by the Merlin Prototyper, the models were converted to the Merode~\cite{snoeck2014overview} format using the Merlin tool~\footnote{\url{https://www.merlin-academic.com/}}. The Merode model is then exported in a XML format as shown in Code~\ref{lst:xml}.

The XML file contains, along with classes (metaobjects) and associations (metadependencies) that represent the class diagram, the list of states (metastate), events (metaevents) and state transitions that represent the state charts that define each class' behaviour. Next, we selected the Merlin Prototyper~\footnote{\url{https://merode.econ.kuleuven.be/pages/CodeGeneration.html}} as code generator because it is capable of generating full functional code from UML class diagrams and state chart and already comes with a built-in Model Driven Testing approach called TesCaV~\cite{marin2020TesCaV}.
\begin{lstlisting}[label=lst:xml,caption=Conceptual Model XML]
<?xml version="1.0" encoding="UTF-8"?>
<mxp:mermaidmodel xmlns:mxp="..." jmermaid.version="1.0">
  
  <mxp:metamodel lastid="116">

    <mxp:metaobjects>
      <mxp:metaobject id="1" name="Accountholder">
        <mxp:metaattributes> ... </mxp:metaattributes>               
        <mxp:metastates> ... <mxp:metastates/> 
      </mxp:metaobject>
    </mxp:metaobjects>
    
    <mxp:metaevents>
      <mxp:metaevent id="104" name="EVcrAccountholder" ../>
    </mxp:metaevents>
    
    <mxp:metadependencies>
      <mxp:metadependency master="1" dependent="57" .../>
    </mxp:metadependencies>
    
    <mxp:metamethods>
      <mxp:metamethod id="8" name="MEcrAccountholder" .../>
    </mxp:metamethods>
  </mxp:metamodel>
  
  <mxp:guimodel currentview="1">
  ...
  </mxp:guimodel>
  
</mxp:mermaidmodel>
\end{lstlisting}
The Merlin Prototyper accepts as input the UML diagrams in XML format generated by the Merlin tool and, using Java templates in Apache Velocity~\footnote{\url{https://velocity.apache.org/}} , composes a fully running application equivalent to the conceptual model that includes a simple JSwing~\footnote{\url{https://openjdk.org/groups/swing/}} user interface for simple testing, as well as a persistence and transaction layer based on the Hibernate\footnote{\url{https://hibernate.org/}} framework, which would be considered \emph{legacy code} nowadays. 

The Merode model, fed into the Merlin Prototyper, generates a Java application, that we now call legacy, that is fully functional and equivalent to the model. The Merode application follows this structure:
\begin{verbatim}
project/merode_application/src
    ├── dao
    ├── tescav
    └── ui
\end{verbatim}

For the sake of brevity, we have omitted some marginal folders, but we will describe the core concepts: the \texttt{dao} folders contains the Data Access Object classes that implement all the classes defined in the model with the proper constraints. The \texttt{tescav} folder contains all test cases defined for the application, while the \texttt{ui} folder contains the files for the JSwing user interface.

We observe that, due to the nature of the template-based generation~\cite{monsieur2006pim}, the application generated by the Merlin Prototyper does not carry any of the business logic, but offers only conformity to the classes, associations and states with transitions. TesCaV offers the possibility to test the generated application in ten different criteria related class to class diagram-based criteria, transition-based criteria and data-based criteria: TesCaV implements test criteria related to class attributes, association-end-multiplicity, and generalization criteria related to the class diagram; It also implements the all-states, all-transitions, all-transitions-pair, all loop-free-paths, all-one-loop-paths, and all-configurations criteria related to the state transition diagram; Finally, TesCaV implements the one-value criterion related to data. 

Every application generated with Merlin Prototyper always passes all TesCaV cases, and we want ideally that the same will remain true for the refactored application. Because of the nature of TesCaV there are no test for the user interface part, hence we will not include in tests the refactored front-end part, while we need to still be fully functional.

We put together the test set by providing for each test case: \emph{i)} the textual description, \emph{ii)} the corresponding conceptual model in xml format, \emph{iii)} the corresponding Merlin-generated Java application and \emph{iv)} a list of all test from TesCaV for that application. The finalized dataset used for our test runs is available at~\url{https://zenodo.org/records/19222552}.

\subsection{\name{} Architecture}
\name{} was designed to be easy to use and to generate reliable code using modern technology stacks. Adding support for new language or technology is as simple as writing a prompt to your favorite AI. We chose to target the refactoring of the monolithic legacy Java application with JSwing user interface created by the Merlin prototyper, outputting a modern web-application with a RESTful backend written in Java with Spring Boot and a front-end written in JavaScript with React framework\footnote{\url{react.dev}} support.\\
We decided to adopt a Multi-Agent architecture~\cite{li2024survey}, where a \emph{agent} is a LLM capable of accessing external tools and performing automatically subsequent action. The access to external tools allow \name{} to directly interact with the legacy code generated by the Merlin Prototyper by reading and writing files and directly extracting needed information. \name{}'s tools also support terminal commands execution for compiling and debugging the application. The other major advantage of the multi-agent paradigm is the \emph{divide et impera} built-in approach for every step of the refactoring: the agents do not attempt to create the whole application in one pass, but rather break down the task into smaller sub-tasks --like identifying the classes to implement or making sure the application will compile. This allows \name{} to be more reliable than simply prompting your favorite AI in the chat. \name{} was built using the OpenHands SDK~\footnote{\url{https://docs.openhands.dev/sdk}} Python library.\\
We decided to build the framework with three LLM main agents: \emph{i)} a Backend Agent, \emph{ii)} a Front-end Agent, \emph{iii)} a Test Execution Agent. Before acting, each of the main agents puts down a to-do list with goals to achieve based on the prompt and is allowed to spawn up to 10 sub-agents to perform simpler tasks off the to-do list. The overall architecture is presented in Figure~\ref{fig:archi}. \\
As the name suggests, each agent will specifically generate a section of the complete architecture. In particular, the test execution agent will extract the tests generated by TesCaV hard-coded into the final Merlin Prototyper output, adapt them for the new application, execute them and write a report with the number of test passed, failed and skipped. \\
The structure of the prompts that add support to different technology stacks and guides the agent is shared with all three agents: first the \emph{task} are given, e.g. creating a spring boot backend from legacy code, then the \emph{background} with a brief description of the task, e.g. the textual description description of the business case; next, the \emph{requirements} are given, e.g. where to find the legacy and what libraries to use for the implementation; then the \emph{constraints} are listed, e.g. to use only the indicated technology stack; and finally, the \emph{success criteria} that once fulfilled will make the agent stop, e.g. the projects compiles without errors.\\
Finally, it also worth noticing that building this powerful architecture was made possible only recently with the new generation of LLMs like Claude 4.6, Gpt-5, Gemini 3 Pro or Kimi 2.5 that not only offer native support to agentic use, but also are sufficiently competent in programming skills~\cite{plaat2025agentic}. As a result, \name{} is powered by one of the most capable LLMs available in the first quarter of 2026: Claude 4.6 Sonnet~\footnote{\url{https://www.anthropic.com/news/claude-sonnet-4-6}}.

\subsection{Testing Metrics}
For testing the functionality of the refactored code, together with test passing rate, time of generation and cost, we chose these metrics very popular in literature for benchmarking code generated by LLMs~\cite{jiang2026survey}:

\begin{center}
\fbox{
    \begin{tabular}{c}
    $\text{compile@k} = 1 - \frac{\binom{n-c}{k}}{\binom{n}{k}}$ \\[10pt]
    $\text{pass@k} = 1 - \frac{\binom{n-c}{k}}{\binom{n}{k}}$ \\[10pt]
    $\text{pass}^k = \left( \frac{c}{n} \right)^k$ 
    \end{tabular}
}
\end{center}

\noindent Where $n$ is the number of tests, $c$ is the number of successful tests, and $k$ is indicates how many generation attempts to evaluate. In our testing, \texttt{compile@k} offers the estimated probability of a project refactored by \name{} compiling after $k$ generation attempts (of the same project); \texttt{pass@k} offers the estimated probability of a project to be 100\% compliant in relation to the TesCaV tests after $k$ generation attempts (of the same project); \texttt{pass$^k$} offers the estimated probability of \emph{all the $k$ generations of a project} to be 100\% compliant in relation to the TesCaV tests after $k$ generation attempts (of a projects).
Those metrics aim at estimating the correctness of a probabilistic generation approach like the ones powered by LLMs, where results of a single run could be not exactly reproducible.\\
However, since the standard \texttt{pass@k} metric computed at a project level does not tell the whole story about the quality of the generation --since it would count as failed even a project with 99\% passing test rate, we computed \texttt{pass@k} for the single unit test from TesCaV, computing the estimated probability of passing a single test in $k$ generation attempt of the same projects, offering a much finer granularity.

For example, a \texttt{compile@1}=50\% is telling us that a single attempt (k=1) at refactoring a project with our tool would would have one in two chances of compiling correctly. At the same time, a value of \texttt{pass@3}=75\% --computed at a project level-- suggest that, in three different attempts (k=3) at refactoring the same project, you would have three in four chances that at least one attempt would pass all tests. While the same metric, computed at a test level, means that under the same premises, you would have three in four chances that you pass one random unit test. Finally, a value of \texttt{pass$^5$}=20\% is telling that, in five different attempts (k=5) at refactoring the same project, you would have only one in five chances that all the five generation of the same project would pass all tests. \\

\subsection{Experimental setup}
Since we have already defined the target programming languages and technology that we want to generate for, to answer RQ1 we use \name{} to make a first round of code improvement, by refactoring each case in the dataset.

To answer RQ2, we use the metrics of \texttt{compile@k}, \texttt{pass@k} and \texttt{pass$^k$} on the full architecture of each case, with $k \in \{1,3,5\}$ where the test cases are the ones defined in the TesCaV test set. Our goal is to obtain an application that, while having modern technologies on full display, still remains capable of passing all the test cases, like the ones generated by the Merlin Prototyper. In other words, the improved code resulting from RQ1 should deliver a modern application that is fully consistent with its original conceptual model and performs at least as well as the original code.

Finally, regarding RQ3, we need to test to what extent adding UML class and state charts impacts the code generation. We want to test two different input cases: \emph{i)} only the code generated by the Merlin Prototyper along with the textual description of the target application, \emph{ii)} the same as \emph{i)} but providing the corresponding UML class and state charts instead of the Merlin-code.

Given this experimental setup, we need to generate 8 refactored application using 2 input types, resulting in 16 generations of front-end and backend, each complemented with a full test cases run to validate the performance of the resulting code.

\subsubsection{The Pilot Run}
\begin{center}
    \begin{figure}[htbp]
        \centering
        \includegraphics[width=\linewidth]{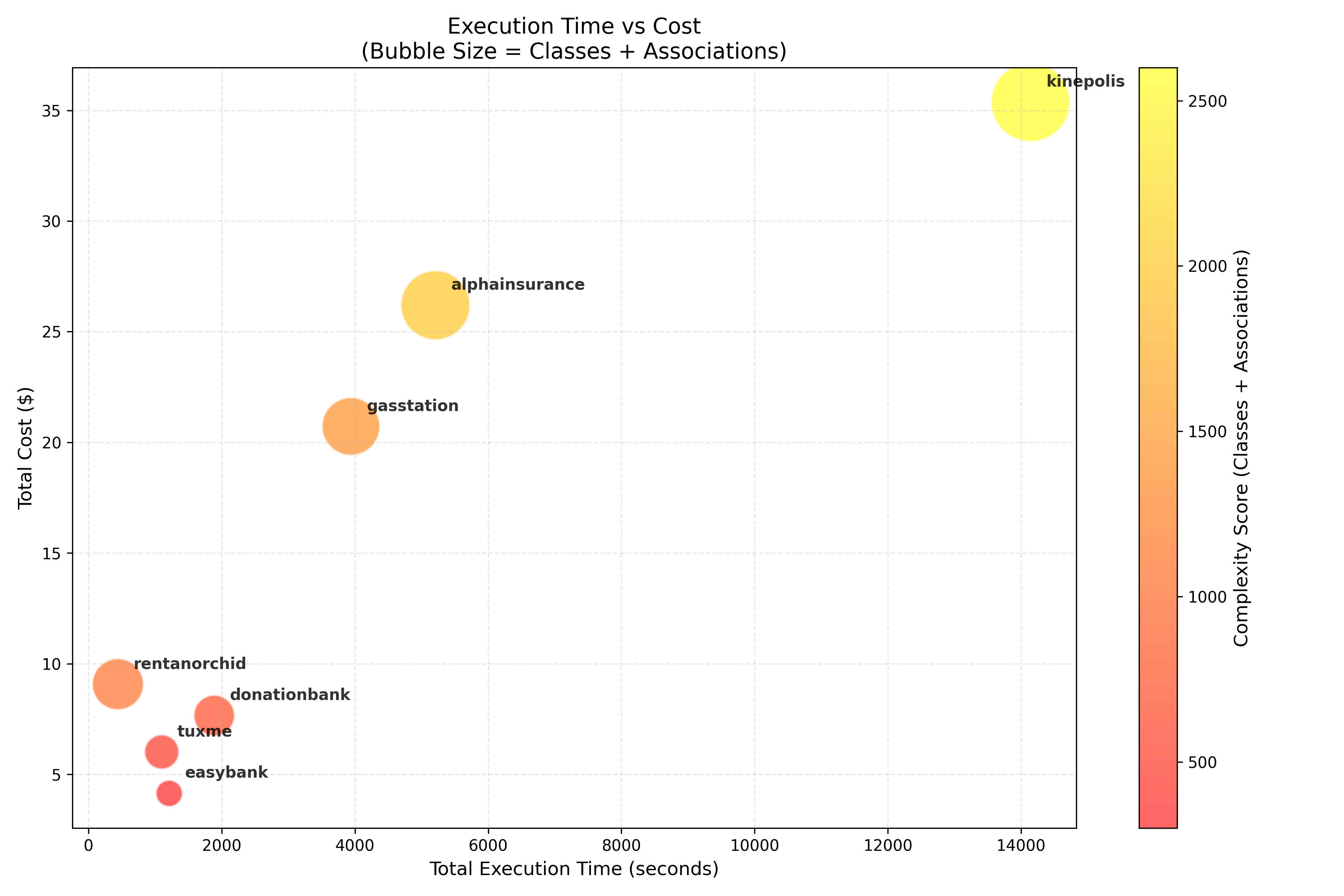}
        \caption{Pilot Run Results in Terms of Cost and Execution Time Broken Down by Cases Complexity}
        \label{fig:t1r}
        \Description{The full path description}
    \end{figure}
\end{center}

Using \name{}'s initial multi-agent architecture we executed a first round of code generation, providing the Merlin-code and the text description as input, and requesting an output making use of Java and Spring Boot for the backend and JavaScript with React for the front-end.
The results in terms of generation time and money, and of the TesCaV passing rate are reported respectively in Table~\ref{tab:updated_project_metrics_sorted} and Table~\ref{tab:test_results}. Despite the almost perfect test results according to TesCaV metrics, the cost in terms of time and money seems to grow more than linearly with respect to the complexity of the cases, as can be observed from Figure~\ref{fig:t1r}. \\

\subsubsection{Adjusted Architecture and Experimental Setup}
Following the encouraging yet expensive results of the pilot run, we decided to optimize our initial naïve approach by adding a more efficient LLM as a base for our agents, to make the architecture more effective in terms of price to performance ratio. In particular, we decided to delegate the execution of the backend and the front-end agents to a less powerful LLM, since the task did not require request particularly demanding, while keeping the test agent powered by Claude 4.6 because the test handling task was deemed more demanding and critical. 

\begin{table*}[t]
    \centering
    
    % First minipage for the 3-column table (takes up 45% of the text width)
    \begin{minipage}{0.46\textwidth}
        \centering
        \caption{Pilot Run Cost Summary}
        \label{tab:updated_project_metrics_sorted}
        \resizebox{\linewidth}{!}{%
            \begin{tabular}{lrr}
            \toprule
            \textbf{Project Name} & \textbf{Total Time (min)} & \textbf{Total Cost (\$)} \\
            \midrule
            alphainsurance & 86.82  & 26.1938 \\
            donationbank   & 31.47  & 7.6568  \\
            easybank       & 20.20  & 4.1375  \\
            gasstation     & 65.78  & 20.7478 \\
            hfte           & 84.39  & 10.0700 \\
            kinepolis      & 235.71 & 35.3700 \\
            rentanorchid   & 40.78  & 9.0976  \\
            tuxme          & 18.34  & 6.0094  \\
            \midrule
            \textbf{Total} & \textbf{583.50} & \textbf{119.2829} \\
            \textbf{Avg. per Project} & \textbf{72.94} & \textbf{14.9104} \\
            \bottomrule
            \end{tabular}%
        }
    \end{minipage}\hfill % \hfill creates flexible spacing between the two minipages
    % Second minipage for the 5-column table (takes up 52% of the text width)
    \begin{minipage}{0.45\textwidth}
        \centering
        \caption{Pilot Run Test Execution Results}
        \label{tab:test_results}
        \resizebox{\linewidth}{!}{%
            \begin{tabular}{lrrrr}
            \toprule
            \textbf{Project Name} & \textbf{Total} & \textbf{Passed} & \textbf{Failed} & \textbf{Pass Rate} \\
            \midrule
            alphainsurance & 127 & 125 & 2 & 98.43\% \\
            donationbank   & 96  & 96  & 0 & 100.00\% \\
            easybank       & 91  & 91  & 0 & 100.00\% \\
            gasstation     & 138 & 138 & 0 & 100.00\% \\
            hfte           & 88  & 88  & 0 & 100.00\% \\
            kinepolis      & 119 & 119 & 0 & 100.00\% \\
            rentanorchid   & 108 & 107 & 1 & 99.07\% \\
            tuxme          & 58  & 57  & 1 & 98.28\% \\
            \midrule
            \textbf{Total} & \textbf{825} & \textbf{821} & \textbf{4} & \textbf{99.52\%} \\
            \textbf{Avg per Proj.} & \textbf{103.13} & \textbf{102.63} & \textbf{0.50} & \textbf{99.47\%} \\
            \bottomrule
            \end{tabular}%
        }
    \end{minipage}

\end{table*}

\begin{center}
    \begin{figure}[H]
        \centering
        \includegraphics[width=\linewidth]{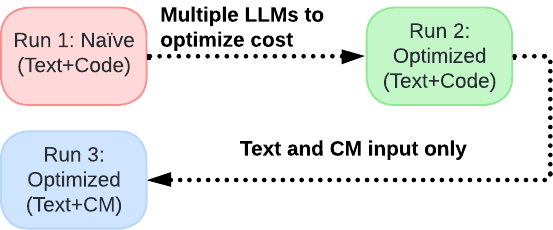}
        \caption{Optimized test path}
        \label{fig:paths}
        \Description{The full test path outlined}
    \end{figure}
\end{center}

The baseline was set with Claude 4.6 Sonnet at 25 USD/Million output token~\footnote{\url{https://platform.claude.com/docs/en/about-claude/pricing}} (as per March 2026), so we decide to test one of the most competitive LLMs in terms of pricing: Mistral AI's Codestral 2 with only 2 USD/Million output token~\footnote{\url{https://docs.mistral.ai/models/devstral-2-25-12}}. However, despite the important price cut, the result with front-end and backend were simply not up to expectations, producing incomplete results (failed run test are also included in the Zenodo). Finally, the sweet spot seemed to be Kimi K2.5\footnote{\url{https://github.com/MoonshotAI/Kimi-K2.5}}, an open-source LLM with a pricing of 3 USD/Million output token\footnote{\url{https://platform.moonshot.ai/docs/pricing/chat\#product-pricing}}.\\
To observe distinctively the impact of the Conceptual Model on the new code generation, we decided to make a test run without the legacy code as input to refactor but only the conceptual model, to see if that is sufficient to achieve satisfactory results.\\
The remainder of the \name{} testing is described in Figure~\ref{fig:paths}: we will run our Optimized mainly Kimi-powered refactoring approach with the same parameters as the first run, and finally we will perform a third run with the Optimized \name{}, having as inputs the conceptual models instead of the project's generated code.

\section{Results} \label{sec:rs}
To evaluate the efficacy of our proposed \name{} architecture, we analyzed the generation runs across the dataset of eight projects, examining execution time, financial cost, and functional reliability metrics. \\
For the clarity of analyzing the full test runs of \name{}, we will from now use the label of Run 1 for referring to the pilot run using text and Merlin-code as input, the label Run 2 for referring to the second run using the optimize architecture and text and Merlin-code as input, and the label Run 3 for referring to the optimized run using the text and conceptual model as input.

\subsection{Optimized Runs: Cost-Benefit Analysis}
\begin{table*}[htbp]
    \centering
    
    % First minipage for the first table (takes up 58% of the text width)
    \begin{minipage}{0.38\textwidth}
    \centering
        \caption{Cost Summary of R2 and R3}
        \label{tab:execution_cost_comparison_sorted}
        \resizebox{\linewidth}{!}{%
            \begin{tabular}{lrr|rr}
            \toprule
            \textbf{Project} & \multicolumn{2}{c}{\textbf{Total Time (min)}} & \multicolumn{2}{c}{\textbf{Total Cost (\$)}} \\
            \cmidrule(lr){2-3} \cmidrule(lr){4-5}
            & \textbf{Run 2} & \textbf{Run 3} & \textbf{Run 2} & \textbf{Run 3} \\
            \midrule
            alphainsurance & 38.91  & 37.21  & 13.2098 & 6.9385  \\
            donationbank   & 59.08  & 48.08  & 4.0238  & 2.3997  \\
            easybank       & 14.73  & 12.97  & 2.5746  & 1.9217  \\
            gasstation     & 85.10  & 67.53  & 9.4993  & 4.3861  \\
            hfte           & 172.10 & 82.74  & 21.4900 & 10.1790 \\
            kinepolis      & 72.87  & 29.91  & 7.0114  & 5.4938  \\
            rentanorchid   & 49.09  & 57.50  & 1.9551  & 3.6545  \\
            tuxme          & 54.94  & 39.44  & 2.6634  & 1.8437  \\
            \midrule
            \textbf{Total} & \textbf{546.82} & \textbf{375.38} & \textbf{62.4274} & \textbf{36.8170} \\
            \textbf{Avg per Proj.} & \textbf{68.35} & \textbf{46.92} & \textbf{7.8034} & \textbf{4.6021} \\
            \bottomrule
            \end{tabular}%
        }
    \end{minipage}\hfill % \hfill pushes the tables to the left and right edges
    % Second minipage for the second table (takes up 38% of the text width)
    \begin{minipage}{0.56\textwidth}
        \centering
        \caption{Comparison of R2 and R3 Test Execution Results}
        \label{tab:combined_runs}
        % Resizebox ensures the wide table scales down to fit the minipage
        \resizebox{\linewidth}{!}{%
            \begin{tabular}{l | rrrr | rrrr}
            \toprule
            \textbf{Project} & \multicolumn{4}{c|}{\textbf{Run 2}} & \multicolumn{4}{c}{\textbf{Run 3}} \\
            \cmidrule(lr){2-5} \cmidrule(lr){6-9}
            & \textbf{Total} & \textbf{Pass} & \textbf{Fail} & \textbf{Rate} & \textbf{Total} & \textbf{Pass} & \textbf{Fail} & \textbf{Rate} \\
            \midrule
            alphainsurance & 300 & 187 & 113 & 62.33\% & 230 & 228 & 2  & 99.13\% \\
            donationbank   & 71  & 71  & 0   & 100.00\% & 88  & 88  & 0  & 100.00\% \\
            easybank       & 64  & 59  & 5   & 92.19\%  & 57  & 57  & 0  & 100.00\% \\
            gasstation     & 236 & 217 & 18  & 91.95\%  & 169 & 141 & 28 & 83.43\% \\
            hfte           & 229 & 46  & 129 & 20.09\%  & 181 & 179 & 2  & 98.90\% \\
            kinepolis      & 179 & 75  & 39  & 41.90\%  & 162 & 156 & 6  & 96.30\% \\
            rentanorchid   & 113 & 113 & 0   & 100.00\% & 156 & 154 & 2  & 98.72\% \\
            tuxme          & 46  & 46  & 0   & 100.00\% & 47  & 47  & 0  & 100.00\% \\
            \midrule
            \textbf{Total} & \textbf{1,238} & \textbf{814} & \textbf{304} & \textbf{65.75\%} & \textbf{1,090} & \textbf{1,050} & \textbf{40} & \textbf{96.33\%} \\
            \textbf{Avg per Proj.} & \textbf{154.75} & \textbf{101.75} & \textbf{38.00} & \textbf{88.56\%} & \textbf{136.25} & \textbf{131.25} & \textbf{5.00} & \textbf{97.06\%} \\
            \bottomrule
            \end{tabular}%
        }
    \end{minipage}

\end{table*}
The optimization of the generation process with the aid of Kimi K2.5 produced mixed result i from what emerges from Table~\ref{tab:execution_cost_comparison_sorted} and Table~\ref{tab:combined_runs}: whilst the cost cutting was successful --Run 2 and 3 moved to a less-than-linear growth of costs in terms of project complexity compared to the quadratic growth seen in Run 1--, the performance in terms of passing tests were mixed.

Table~\ref{tab:execution_cost_comparison_sorted} highlights a significant disparity in test execution results between the two inputs. Run 2 (Optimized with Merlin Code as input) struggled with functional accuracy, achieving an overall pass rate of only 65.75\%, with projects like \textit{hfte} and \textit{kinepolis} dropping to pass rates of 20.09\% and 41.90\%, respectively. In contrast, Run 3 (Optimized with Conceptual Models as input) demonstrated a massive recovery in quality, achieving an overall pass rate of 96.33\% and securing a 100.00\% pass rate in three projects.

\begin{center}
    \begin{figure*}[htbp]
        \centering
        \includegraphics[width=\linewidth]{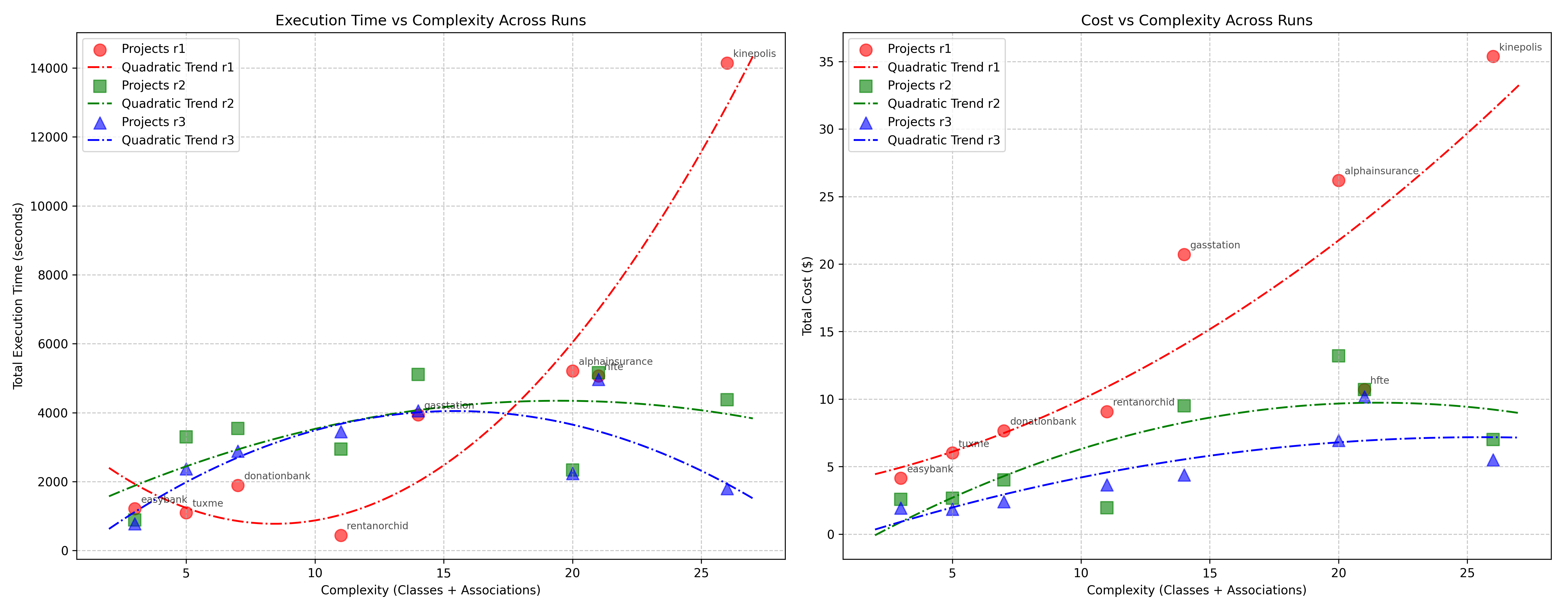}
        \caption{Interpolation Trend Difference Between R1, R2 and R3}
        \label{fig:interpolationdiff}
        \Description{The full path description}
    \end{figure*}
\end{center}

Table~\ref{tab:combined_runs} and Figure~\ref{fig:interpolationdiff} contrast the execution metrics of these two optimized runs. Run 2 incurred a higher total financial cost (\$62.42) compared to Run 3 (\$36.81). Run 3 was also substantially faster, requiring only 375.38 minutes in total, compared to the 546.82 minutes consumed by Run 2. This indicates that supplying the multi-agent system with structured conceptual models rather than legacy code allows it to synthesize the target architecture much faster, producing surprisingly better results.

\subsection{Reliability Metrics For The Three Runs}

\begin{sloppypar}
To rigorously answer how the generated code behaves under modern LLM evaluation standards, we calculated the \texttt{compile@k}, \texttt{pass@k}, and consistency metric $pass^{k}$ for $k \in \{1, 3, 5\}$.
\begin{table*}[htbp]
\centering
\caption{Application-Level Reliability: Pass@k, Pass$^k$, and Compile@k Metrics}
\label{tab:pass_at_k_vs_power_k_100comp}
\small
\begin{tabular}{lc | ccc | ccc | ccc}
\toprule
\textbf{Run} & \textbf{c} & \multicolumn{3}{c|}{\textbf{Pass@k}} & \multicolumn{3}{c|}{\textbf{Pass$^k$}} & \multicolumn{3}{c}{\textbf{Compile@k}} \\
\cmidrule(lr){3-5} \cmidrule(lr){6-8} \cmidrule(lr){9-11}
& & \textbf{k=1} & \textbf{k=3} & \textbf{k=5} & \textbf{k=1} & \textbf{k=3} & \textbf{k=5} & \textbf{k=1} & \textbf{k=3} & \textbf{k=5} \\
\midrule
Run 1 (Pilot)   & 5 & 62.50\% & 98.21\% & 100.00\% & 62.50\% & 24.41\% & 9.54\% & 100.00\% & 100.00\% & 100.00\% \\
Run 2 (Optimized) & 3 & 37.50\% & 82.14\% & 98.21\%  & 37.50\% & 5.27\%  & 0.74\% & 100.00\% & 100.00\% & 100.00\% \\
Run 3 (CM Only)    & 3 & 37.50\% & 82.14\% & 98.21\%  & 37.50\% & 5.27\%  & 0.74\% & 100.00\% & 100.00\% & 100.00\% \\
\bottomrule
\end{tabular}
\end{table*}

\begin{table*}[htbp]
\centering
\caption{Test-Level Reliability: Pass@k, Pass$^k$, and Compile@k Metrics}
\label{tab:test_level_reliability_100comp}
\small
\begin{tabular}{lrr | ccc | ccc | ccc}
\toprule
\textbf{Run} & \textbf{n} & \textbf{c} & \multicolumn{3}{c|}{\textbf{Pass@k}} & \multicolumn{3}{c|}{\textbf{Pass$^k$}} & \multicolumn{3}{c}{\textbf{Compile@k}} \\
\cmidrule(lr){4-6} \cmidrule(lr){7-9} \cmidrule(lr){10-12}
& & & \textbf{k=1} & \textbf{k=3} & \textbf{k=5} & \textbf{k=1} & \textbf{k=3} & \textbf{k=5} & \textbf{k=1} & \textbf{k=3} & \textbf{k=5} \\
\midrule
Run 1 (Pilot) & 825   & 821   & 99.52\% & 100.00\% & 100.00\% & 99.52\% & 98.56\% & 97.61\% & 100.00\% & 100.00\% & 100.00\% \\
Run 2 (Optimized)& 1,238 & 814   & 65.75\% & 96.00\%  & 99.53\%  & 65.75\% & 28.42\% & 12.29\% & 100.00\% & 100.00\% & 100.00\% \\
Run 3  (CM Only)& 1,090 & 1,050 & 96.33\% & 99.99\%  & 100.00\% & 96.33\% & 89.39\% & 82.94\% & 100.00\% & 100.00\% & 100.00\% \\
\bottomrule
\end{tabular}
\end{table*}
 As presented in Table~\ref{tab:pass_at_k_vs_power_k_100comp} and Table~\ref{tab:test_level_reliability_100comp}, all three runs demonstrated flawless syntactical generation, maintaining a $Compile@k$ of 100.00\% across all values of $k$. However, the $Pass@k$ and consistency metrics varied significantly. The optimized model utilizing legacy code (Run 2) exhibited a poor $Pass^{5}$ consistency of only 0.74\% at the project level, reflecting high unreliability across repeated generation attempts. Run 3, guided by conceptual models, restored these reliability metrics to match the Naïve approach, achieving a $Pass@1$ (At least 1) of 37.50\% and matching Run 1's consistency scores perfectly at the project level. At the individual test level (Table 7), Run 3 achieved an impressive $Pass@5$ of 100.00\% and a $Pass^{5}$ consistency of 82.94\%, decisively proving the added value of conceptual models in the generation pipeline.
 \subsection{Test Failures Broken Down By Category}
We were also interested in finding out if across the runs there are some categories of TesCaV tests where \name{} is consistently failing. We decided to aggregate in Table~\ref{tab:testcav_results} all test run results together since the failure pattern across run was similar.
\begin{table}[htbp]
\centering
\caption{Test Results by Criterion}
\label{tab:testcav_results}
\resizebox{\columnwidth}{!}{%
\begin{tabular}{lrrrrr}
\toprule
\textbf{Criterion} & \textbf{Pass} & \textbf{Fail} & \textbf{Skip} & \textbf{Tot} & \textbf{Rate} \\
\midrule
Association-End-Multiplicity           & 131 & 25 & 10 & 166 & 79\% \\
All-Transition-Pairs                    & 220 & 48 & 7  & 275 & 80\% \\
All-Loop-Free-Paths                     & 225 & 42 & 6  & 273 & 82\% \\
All-One-Loop-Paths                      & 178 & 37 & 3  & 218 & 82\% \\
All-Loops                               & 127 & 24 & 4  & 155 & 82\% \\
All-Transitions                         & 647 & 62 & 48 & 757 & 85\% \\
All-Methods                             & 515 & 60 & 21 & 596 & 86\% \\
All-States                              & 309 & 39 & 8  & 356 & 87\% \\
End Objects                             & 160 & 17 & 4  & 181 & 88\% \\
Create Objects                          & 166 & 10 & 8  & 184 & 90\% \\
Method Allowed/Disallowed               & 89  & 10 & 0  & 99  & 90\% \\
Generalization                          & 94  & 0  & 1  & 95  & 99\% \\
\bottomrule
\end{tabular}
}
\end{table}

We can observe that the categories where the highest percentage of errors occurs are about Association end multiplicity (defines the number of entity type instances that can be at one end of an association) and all tests about finite state transitions pairs and loop. We will offer our vision of why that could be the case in the discussion section.

\end{sloppypar}

\section{Discussion}\label{sec:ds}
In this section we will answer to the research question by focusing on key discussion points.
\subsection{RQ1: Results in terms of time and money spent}
The quick answer to RQ1 is: "yes". The current generation of LLMs is capable of powering complex agents specialized in refactoring legacy code by adding support for new technology stacks and different programming languages, without compromising the requirements elicited in the conceptual models. However, if we change the focus to how convenient it is to use \name{}, the answer becomes: "it depends". First of all, we have observed that changing the LLM powering the agents heavily influences the outcome and even the success of the generation of new code. In our tests, we have observed that smaller LLMs like Codestral 2 are still not capable of fully handling the complexity of the task. At the same time, flagship LLMs like Claude 4.6 Sonnet handle the job almost perfectly, but become hefty in terms of time and money required to complete the generation. 

Finding the sweet spot becomes a matter of meeting the user's expectations: from the perspective of a professional developer user, spending one hour and 15 USD for generating an incomplete but fully functional project is almost like a dream come true. From the perspective of a software engineering student the cost quickly adds up. We, as the authors of this article, have found our balance in terms of cost and results with a hybrid solution of Claude 4.6 Sonnet for handling the testing and Kimi K2.5 for the code generation, even if the \texttt{pass@k} scores are lower. 

The other observation that we can take away if we break down the test results, is that the area where even more powerful LLMs struggle is the handling of association multiplicity and state transition. It has been observed in literature~\cite{calamo2025assessing}, that also in generating UML diagrams LLMs struggle with multiplicity and association the most. From this we can derive that, despite all progress, LLMs still have not complete \emph{understanding} of the association dynamics, even if guided by correct conceptual models or code.

\subsection{RQ2: Metric Results compared with existing literature}
The short answer once again is: "pretty good, in a limited sample testing", but, to answer RQ2 properly, we need to contextualize the performance of \name{} with respect to the existing literature. In particular, since we have tested the \texttt{pass@k}  metrics both in terms of full application compliance and single test compliance, we will compare our results first with iEcoreGen~\cite{he2025hybrid}, which measured \texttt{pass@k} and \texttt{compile@k} metrics with respect to the full application; and second with~\cite{wan2025structure} and~\cite{han2024archcode}, which tested their approach on a single function or task, defined in a dataset like HumanEval. Because of being made across different datasets, this kind of comparison is not intended as strict, but rather to position~\name{} in the flourishing landscape of AI-powered MDE tools.\\
The best performance obtained by iEcoreGen in ~\cite{he2025hybrid}, are \texttt{pass@1}=75\% and \texttt{pass@3}=89\%, while \texttt{compile@1}=100\% and \texttt{compile@3}=100\%. \name{} matched the perfect score on \texttt{compile@k} metrics, falls comparatively short in \texttt{pass@1} for all three runs, but achieved better results in the first run in the \texttt{pass@3} metric. %The important difference of at least ten percent in the best case in \texttt{pass@1} metric can be partially explained with the difference of test cases per each project: our average test number is 100 per project whilst around 30 for iEcoreGen~\cite{he2025hybrid}. 
The performance gap of 10\% (compared to our test run) and more can be partially explained by the different testing approach: as TesCaV ensures full model coverage, the average number of tests per case is more than three times higher (100 compared to 30), increasing the probability of failing a test. On the other hand, the smaller dataset of 8 cases compared to the 20 cases in iEcoreGen also contributes to penalizing our results as not a single app passed 100\% of the test cases.\\
Since we felt that the pass metric in at application level was only partially describing  \name{}'s capabilities, we decided also to measure the \texttt{pass@k} for the single test case. This approach was also followed in ~\cite{wan2025structure}, where \texttt{pass@1} was around 95\%, and in ARCHCODE ~\cite{han2024archcode}, where respectively \texttt{pass@1},3,5= 86.36\% 88.62\% 90.48\%. These more refined numbers suggest that Run 1 results were above the current literature, Run 2 result, despite the low \texttt{\texttt{pass@1}} were better as well, but more interestingly, Run 3 results, the ones without using the legacy code as input, were better as well. The explanation for this counterintuitive result, where conceptual model-only code generation was more successful than legacy code refactoring, has to be found in the structure of the conceptual model. We will discuss this in detail the next section.\\ 
A relevant metric not found in literature is \texttt{pass$^k$}. \name{} scores quite low for this application level metric, suggesting that despite the incredible capabilities of LLMs, we are still far from generating a perfectly compliant application every time, expecially with less capable LLMs.

\subsection{RQ3: Adding Conceptual Models}
A quantitative answer to RQ3 could be: "using the conceptual model as input to \name{} led to about 10\% more test passed per project in average", but we need to elaborate it more. The most interesting takeaway from our experiment runs is that even with the less powerful LLMs powering \name{} and with the input of the conceptual model without the Merlin generated code, we were able to achieve similar and in some cases even better results than the Run 2 that could use the full legacy code base as a reference. This result can be partially explained by the XML format of the conceptual models (See Code~\ref{lst:xml}): the XML file clearly states the various part that are going to be tested with TesCaV, like attributes, object states, events and methods. But if the legacy code is equivalent to the model how come the TesCaV results are so different between Run 2 and 3? Our empirical explanation is that, while Merlin-generated code still has all the elements necessary to pass tests, they are diluted in a vast legacy code base, and thus the less powerful llms like Codestral or even Kimi K2.5 do not manage to capture all those elements perfectly.

These results suggest that conceptual modeling could still be crucial in a faster evolving coding landscape where the speed and apparently flawless generation of LLMs are dominant. Our experiments proved that the use of a conceptual model could be more effective for generating code compliant to the specifications instead of prompting it to refactor the legacy code base, as can be done in tools such as Visual Studio Code or Cursor.
\subsection{Limitations and threats to validity}
While the \name{} framework demonstrates promising results, several limitations and threats to validity must be acknowledged. First, it is necessary to acknowledge that the agentic loop after having observed too much test failures attempts to fix the code and perform again a sequential testing phase. This paper only reports only the final results, as the intermediate test results could underestimate the capability of the framework that showed an unexpected positive behavior. However, while this iterative refinement improves the final output, it introduces the risk that the agent might over-optimize for specific test cases rather than addressing broader architectural flaws. Second, we observed slightly different numbers of unit tests for a single case across our evaluations due to differences in the generated architecture. Transitioning from a monolithic Java legacy system to a decoupled Spring Boot and React stack inherently alters the application's structure, making a perfect one-to-one mapping of unit tests challenging and potentially skewing comparative pass rates. Finally, the success of the generation pipeline is heavily dependent on the input quality; a well-done conceptual model is needed to achieve high reliability. Furthermore, our current evaluation is limited because we have a small set of test cases. Validating the framework on only eight projects restricts the immediate generalizability of our findings to more extensive, industrial-scale applications.

\section{Conclusion and Future Work} \label{sec:fw}
In conclusion, this paper introduces \name{}, a novel multi-agent framework that successfully bridges the gap between Model-Driven Engineering (MDE) and Large Language Models (LLMs). By leveraging conceptual models and legacy code, ARTHUR can refactor outdated architectures into modern technology stacks while maintaining almost complete alignment with original system requirements. Our findings indicate that providing LLM agents with rigorous conceptual models significantly enhances the reliability, cost-effectiveness, and speed of code generation compared to relying on legacy code alone.

To address current limitations and further validate our approach, our future work will focus on two primary directions. First, we plan to evaluate \name{} on a bigger dataset. Expanding our testing to encompass a wider array of complex, real-world MDE projects will help solidify our confidence in the framework's scalability and robustness. Second, we aim to extend \name{}'s agentic capabilities to support Python/Flask or other languages and frameworks. Because the architecture is prompt-driven, adding support for alternative backend and front-end stacks will demonstrate its true language-agnostic potential in modern software engineering.

%% Static bibliography: compile with pdflatex only (no BibTeX required).
% Static bibliography generated from biblio.bib with ACM-Reference-Format.
% This file is included directly by main.tex, so BibTeX is not required.


\begin{thebibliography}{31}

%%% ====================================================================
%%% NOTE TO THE USER: you can override these defaults by providing
%%% customized versions of any of these macros before the \bibliography
%%% command.  Each of them MUST provide its own final punctuation,
%%% except for \shownote{} and \showURL{}.  The latter two
%%% do not use final punctuation, in order to avoid confusing it with
%%% the Web address.
%%%
%%% To suppress output of a particular field, define its macro to expand
%%% to an empty string, or better, \unskip, like this:
%%%
%%% \newcommand{\showURL}[1]{\unskip}   % LaTeX syntax
%%%
%%% \def \showURL #1{\unskip}           % plain TeX syntax
%%%
%%% ====================================================================

\ifx \showCODEN    \undefined \def \showCODEN     #1{\unskip}     \fi
\ifx \showISBNx    \undefined \def \showISBNx     #1{\unskip}     \fi
\ifx \showISBNxiii \undefined \def \showISBNxiii  #1{\unskip}     \fi
\ifx \showISSN     \undefined \def \showISSN      #1{\unskip}     \fi
\ifx \showLCCN     \undefined \def \showLCCN      #1{\unskip}     \fi
\ifx \shownote     \undefined \def \shownote      #1{#1}          \fi
\ifx \showarticletitle \undefined \def \showarticletitle #1{#1}   \fi
\ifx \showURL      \undefined \def \showURL       {\relax}        \fi
% The following commands are used for tagged output and should be
% invisible to TeX
\providecommand\bibfield[2]{#2}
\providecommand\bibinfo[2]{#2}
\providecommand\natexlab[1]{#1}
\providecommand\showeprint[2][]{arXiv:#2}

\bibitem[Buchmann et~al\mbox{.}(2024)]%
        {buchmann2024white}
\bibfield{author}{\bibinfo{person}{Thomas Buchmann}, \bibinfo{person}{Ren{\'e}
  Peinl}, {and} \bibinfo{person}{Felix Schw{\"a}gerl}.}
  \bibinfo{year}{2024}\natexlab{}.
\newblock \showarticletitle{White-box LLM-supported low-code engineering: a
  vision and first insights}. In \bibinfo{booktitle}{\emph{Proceedings of the
  ACM/IEEE 27th International Conference on Model Driven Engineering Languages
  and Systems}}. \bibinfo{pages}{556--560}.
\newblock


\bibitem[Buchmann et~al\mbox{.}(2025)]%
        {buchmann2025model}
\bibfield{author}{\bibinfo{person}{Thomas Buchmann}, \bibinfo{person}{Felix
  Schw{\"a}gerl}, {and} \bibinfo{person}{Ren{\'e} Peinl}.}
  \bibinfo{year}{2025}\natexlab{}.
\newblock \showarticletitle{To Model, to Prompt, or to Code? The Choice Is
  Yours: A Multi-Paradigmatic Approach to Software Development}.
\newblock  (\bibinfo{year}{2025}).
\newblock


\bibitem[Calamo et~al\mbox{.}(2025)]%
        {calamo2025assessing}
\bibfield{author}{\bibinfo{person}{Marco Calamo}, \bibinfo{person}{Massimo
  Mecella}, {and} \bibinfo{person}{Monique Snoeck}.}
  \bibinfo{year}{2025}\natexlab{}.
\newblock \showarticletitle{Assessing the suitability of large language models
  in generating uml class diagrams as conceptual models}. In
  \bibinfo{booktitle}{\emph{International Conference on Business Process
  Modeling, Development and Support}}. Springer, \bibinfo{pages}{211--226}.
\newblock


\bibitem[Chen et~al\mbox{.}(2021)]%
        {chen2021evaluating}
\bibfield{author}{\bibinfo{person}{Mark Chen}, \bibinfo{person}{Jerry Tworek},
  \bibinfo{person}{Heewoo Jun}, \bibinfo{person}{Qiming Yuan},
  \bibinfo{person}{Henrique Ponde De~Oliveira Pinto}, \bibinfo{person}{Jared
  Kaplan}, \bibinfo{person}{Harri Edwards}, \bibinfo{person}{Yuri Burda},
  \bibinfo{person}{Nicholas Joseph}, \bibinfo{person}{Greg Brockman},
  {et~al\mbox{.}}} \bibinfo{year}{2021}\natexlab{}.
\newblock \showarticletitle{Evaluating large language models trained on code}.
\newblock \bibinfo{journal}{\emph{arXiv preprint arXiv:2107.03374}}
  (\bibinfo{year}{2021}).
\newblock


\bibitem[Cheon(2025)]%
        {cheon2025llms}
\bibfield{author}{\bibinfo{person}{Yoonsik Cheon}.}
  \bibinfo{year}{2025}\natexlab{}.
\newblock \showarticletitle{LLMs as Code Generators for Model-Driven
  Development}. In \bibinfo{booktitle}{\emph{Proceedings of the 20th
  International Conference on Software Technologies (ICSOFT)}}.
  \bibinfo{pages}{386--393}.
\newblock


\bibitem[Escobar et~al\mbox{.}(2025)]%
        {escobar2025prompt}
\bibfield{author}{\bibinfo{person}{Miguel Escobar}, \bibinfo{person}{Rene
  Noel}, {and} \bibinfo{person}{Oscar Pastor}.}
  \bibinfo{year}{2025}\natexlab{}.
\newblock \showarticletitle{Prompt-Guided Evaluation of GPT-4o, Gemini and
  DeepSeek on UML-to-Java Code Generation}. In
  \bibinfo{booktitle}{\emph{International Conference on Conceptual Modeling}}.
  Springer, \bibinfo{pages}{210--224}.
\newblock


\bibitem[Fowler(2018)]%
        {fowler2018uml}
\bibfield{author}{\bibinfo{person}{Martin Fowler}.}
  \bibinfo{year}{2018}\natexlab{}.
\newblock \bibinfo{booktitle}{\emph{UML distilled: a brief guide to the
  standard object modeling language}}.
\newblock \bibinfo{publisher}{Addison-Wesley Professional}.
\newblock


\bibitem[Han et~al\mbox{.}(2024)]%
        {han2024archcode}
\bibfield{author}{\bibinfo{person}{Hojae Han}, \bibinfo{person}{Jaejin Kim},
  \bibinfo{person}{Jaeseok Yoo}, \bibinfo{person}{Youngwon Lee}, {and}
  \bibinfo{person}{Seung~Won Hwang}.} \bibinfo{year}{2024}\natexlab{}.
\newblock \showarticletitle{ARCHCODE: Incorporating Software Requirements in
  Code Generation with Large Language Models}. In
  \bibinfo{booktitle}{\emph{62nd Annual Meeting of the Association for
  Computational Linguistics, ACL 2024}}. Association for Computational
  Linguistics (ACL), \bibinfo{pages}{13520--13552}.
\newblock


\bibitem[He et~al\mbox{.}(2025)]%
        {he2025hybrid}
\bibfield{author}{\bibinfo{person}{Xiao He}, \bibinfo{person}{Ru Chen},
  \bibinfo{person}{Zeqing Zhang}, \bibinfo{person}{Yanling Wang}, {and}
  \bibinfo{person}{Qiuyan Dong}.} \bibinfo{year}{2025}\natexlab{}.
\newblock \showarticletitle{A Hybrid Approach for EMF Code Generation: Code
  Templates Meet Large Language Models}.
\newblock \bibinfo{journal}{\emph{arXiv preprint arXiv:2512.05498}}
  (\bibinfo{year}{2025}).
\newblock


\bibitem[Hurst et~al\mbox{.}(2024)]%
        {hurst2024gpt}
\bibfield{author}{\bibinfo{person}{Aaron Hurst}, \bibinfo{person}{Adam Lerer},
  \bibinfo{person}{Adam~P Goucher}, \bibinfo{person}{Adam Perelman},
  \bibinfo{person}{Aditya Ramesh}, \bibinfo{person}{Aidan Clark},
  \bibinfo{person}{AJ Ostrow}, \bibinfo{person}{Akila Welihinda},
  \bibinfo{person}{Alan Hayes}, \bibinfo{person}{Alec Radford},
  {et~al\mbox{.}}} \bibinfo{year}{2024}\natexlab{}.
\newblock \showarticletitle{Gpt-4o system card}.
\newblock \bibinfo{journal}{\emph{arXiv preprint arXiv:2410.21276}}
  (\bibinfo{year}{2024}).
\newblock


\bibitem[Hutchinson et~al\mbox{.}(2011)]%
        {hutchinson2011model}
\bibfield{author}{\bibinfo{person}{John Hutchinson}, \bibinfo{person}{Mark
  Rouncefield}, {and} \bibinfo{person}{Jon Whittle}.}
  \bibinfo{year}{2011}\natexlab{}.
\newblock \showarticletitle{Model-driven engineering practices in industry}. In
  \bibinfo{booktitle}{\emph{Proceedings of the 33rd International Conference on
  Software Engineering}}. \bibinfo{pages}{633--642}.
\newblock


\bibitem[Jiang et~al\mbox{.}(2026)]%
        {jiang2026survey}
\bibfield{author}{\bibinfo{person}{Juyong Jiang}, \bibinfo{person}{Fan Wang},
  \bibinfo{person}{Jiasi Shen}, \bibinfo{person}{Sungju Kim}, {and}
  \bibinfo{person}{Sunghun Kim}.} \bibinfo{year}{2026}\natexlab{}.
\newblock \showarticletitle{A survey on large language models for code
  generation}.
\newblock \bibinfo{journal}{\emph{ACM Transactions on Software Engineering and
  Methodology}} \bibinfo{volume}{35}, \bibinfo{number}{2}
  (\bibinfo{year}{2026}), \bibinfo{pages}{1--72}.
\newblock


\bibitem[Jin et~al\mbox{.}(2024)]%
        {jin2024llms}
\bibfield{author}{\bibinfo{person}{Haolin Jin}, \bibinfo{person}{Linghan
  Huang}, \bibinfo{person}{Haipeng Cai}, \bibinfo{person}{Jun Yan},
  \bibinfo{person}{Bo Li}, {and} \bibinfo{person}{Huaming Chen}.}
  \bibinfo{year}{2024}\natexlab{}.
\newblock \showarticletitle{From llms to llm-based agents for software
  engineering: A survey of current, challenges and future}.
\newblock \bibinfo{journal}{\emph{arXiv preprint arXiv:2408.02479}}
  (\bibinfo{year}{2024}).
\newblock


\bibitem[Johri and Bansal(2018)]%
        {johri2018identifying}
\bibfield{author}{\bibinfo{person}{Vishal Johri} {and}
  \bibinfo{person}{Srividya Bansal}.} \bibinfo{year}{2018}\natexlab{}.
\newblock \showarticletitle{Identifying trends in technologies and programming
  languages using topic modeling}. In \bibinfo{booktitle}{\emph{2018 IEEE 12th
  International Conference on Semantic Computing (ICSC)}}. IEEE,
  \bibinfo{pages}{391--396}.
\newblock


\bibitem[Kent(2002)]%
        {kent2002model}
\bibfield{author}{\bibinfo{person}{Stuart Kent}.}
  \bibinfo{year}{2002}\natexlab{}.
\newblock \showarticletitle{Model driven engineering}. In
  \bibinfo{booktitle}{\emph{International conference on integrated formal
  methods}}. Springer, \bibinfo{pages}{286--298}.
\newblock


\bibitem[Li et~al\mbox{.}(2024)]%
        {li2024survey}
\bibfield{author}{\bibinfo{person}{Xinyi Li}, \bibinfo{person}{Sai Wang},
  \bibinfo{person}{Siqi Zeng}, \bibinfo{person}{Yu Wu}, {and}
  \bibinfo{person}{Yi Yang}.} \bibinfo{year}{2024}\natexlab{}.
\newblock \showarticletitle{A survey on LLM-based multi-agent systems:
  workflow, infrastructure, and challenges}.
\newblock \bibinfo{journal}{\emph{Vicinagearth}} \bibinfo{volume}{1},
  \bibinfo{number}{1} (\bibinfo{year}{2024}), \bibinfo{pages}{9}.
\newblock


\bibitem[Mar{\'\i}n et~al\mbox{.}(2020)]%
        {marin2020TesCaV}
\bibfield{author}{\bibinfo{person}{Beatriz Mar{\'\i}n},
  \bibinfo{person}{Sof{\'\i}a Alarc{\'o}n}, \bibinfo{person}{Giovanni
  Giachetti}, {and} \bibinfo{person}{Monique Snoeck}.}
  \bibinfo{year}{2020}\natexlab{}.
\newblock \showarticletitle{Tescav: An approach for learning model-based
  testing and coverage in practice}. In \bibinfo{booktitle}{\emph{International
  Conference on Research Challenges in Information Science}}. Springer,
  \bibinfo{pages}{302--317}.
\newblock


\bibitem[Monsieur et~al\mbox{.}(2006)]%
        {monsieur2006pim}
\bibfield{author}{\bibinfo{person}{Geert Monsieur}, \bibinfo{person}{Monique
  Snoeck}, \bibinfo{person}{Raf Haesen}, \bibinfo{person}{Wilfried Lemahieu},
  \bibinfo{person}{J Almeida}, \bibinfo{person}{L Pires}, {and}
  \bibinfo{person}{M Van~Sinderen}.} \bibinfo{year}{2006}\natexlab{}.
\newblock \showarticletitle{PIM to PSM transformations for an event driven
  architecture in an educational tool}. In
  \bibinfo{booktitle}{\emph{Proceedings of the European Workshop on Milestones,
  Models and Mappings for Model-Driven Architecture (3M4MDA),}}. University of
  Twente; Twente, Nederland, \bibinfo{pages}{49--64}.
\newblock


\bibitem[Mussa et~al\mbox{.}(2009)]%
        {mussa2009survey}
\bibfield{author}{\bibinfo{person}{Mohamed Mussa}, \bibinfo{person}{Samir
  Ouchani}, \bibinfo{person}{Waseem Al~Sammane}, {and}
  \bibinfo{person}{Abdelwahab Hamou-Lhadj}.} \bibinfo{year}{2009}\natexlab{}.
\newblock \showarticletitle{A survey of model-driven testing techniques}. In
  \bibinfo{booktitle}{\emph{2009 Ninth International Conference on Quality
  Software}}. IEEE, \bibinfo{pages}{167--172}.
\newblock


\bibitem[Plaat et~al\mbox{.}(2025)]%
        {plaat2025agentic}
\bibfield{author}{\bibinfo{person}{Aske Plaat}, \bibinfo{person}{Max van
  Duijn}, \bibinfo{person}{Niki Van~Stein}, \bibinfo{person}{Mike Preuss},
  \bibinfo{person}{Peter van~der Putten}, {and} \bibinfo{person}{Kees~Joost
  Batenburg}.} \bibinfo{year}{2025}\natexlab{}.
\newblock \showarticletitle{Agentic large language models, a survey}.
\newblock \bibinfo{journal}{\emph{Journal of Artificial Intelligence Research}}
   \bibinfo{volume}{84} (\bibinfo{year}{2025}).
\newblock


\bibitem[Romeo et~al\mbox{.}(2025)]%
        {romeo2025uml}
\bibfield{author}{\bibinfo{person}{Joseph Romeo}, \bibinfo{person}{Marco
  Raglianti}, \bibinfo{person}{Csaba Nagy}, {and} \bibinfo{person}{Michele
  Lanza}.} \bibinfo{year}{2025}\natexlab{}.
\newblock \showarticletitle{UML is back. Or is it? Investigating the past,
  present, and future of UML in open source software}. In
  \bibinfo{booktitle}{\emph{2025 IEEE/ACM 47th International Conference on
  Software Engineering (ICSE)}}. IEEE, \bibinfo{pages}{2342--2354}.
\newblock


\bibitem[Sadik et~al\mbox{.}(2024)]%
        {sadik2024llm}
\bibfield{author}{\bibinfo{person}{Ahmed~R Sadik}, \bibinfo{person}{Sebastian
  Brulin}, \bibinfo{person}{Markus Olhofer}, \bibinfo{person}{Antonello
  Ceravola}, {and} \bibinfo{person}{Frank Joublin}.}
  \bibinfo{year}{2024}\natexlab{}.
\newblock \showarticletitle{LLM as a code generator in Agile Model Driven
  Development}. In \bibinfo{booktitle}{\emph{International Conference on
  Model-Driven Engineering and Software Development}}. Springer,
  \bibinfo{pages}{198--212}.
\newblock


\bibitem[Santos et~al\mbox{.}(2025)]%
        {santos2025applying}
\bibfield{author}{\bibinfo{person}{Gon{\c{c}}alo Santos},
  \bibinfo{person}{Clara Silveira}, \bibinfo{person}{Vitor Santos},
  \bibinfo{person}{Arnaldo Santos}, {and} \bibinfo{person}{Henrique Mamede}.}
  \bibinfo{year}{2025}\natexlab{}.
\newblock \showarticletitle{Applying Large Language Models to Software
  Development: Enhancing Requirements, Design and Code}. In
  \bibinfo{booktitle}{\emph{International Conference on Disruptive
  Technologies, Tech Ethics and Artificial Intelligence}}. Springer,
  \bibinfo{pages}{226--239}.
\newblock


\bibitem[Snoeck(2014)]%
        {snoeck2014overview}
\bibfield{author}{\bibinfo{person}{Monique Snoeck}.}
  \bibinfo{year}{2014}\natexlab{}.
\newblock \showarticletitle{Overview of MERODE}.
\newblock In \bibinfo{booktitle}{\emph{Enterprise Information Systems
  Engineering: The MERODE Approach}}. \bibinfo{publisher}{Springer},
  \bibinfo{pages}{51--75}.
\newblock


\bibitem[Verbruggen et~al\mbox{.}(2025)]%
        {verbruggen2025toward}
\bibfield{author}{\bibinfo{person}{Charlotte Verbruggen},
  \bibinfo{person}{Lukas Netz}, \bibinfo{person}{Philipp-Lorenz Glaser},
  \bibinfo{person}{Marion Scholz}, \bibinfo{person}{Christian Huemer},
  \bibinfo{person}{Marco Calamo}, \bibinfo{person}{Bernhard Rumpe},
  \bibinfo{person}{Monique Snoeck}, {and} \bibinfo{person}{Dominik Bork}.}
  \bibinfo{year}{2025}\natexlab{}.
\newblock \showarticletitle{Toward a Community-Curated Golden Dataset of UML
  Models}. In \bibinfo{booktitle}{\emph{2025 ACM/IEEE 28th International
  Conference on Model Driven Engineering Languages and Systems Companion
  (MODELS-C)}}. IEEE, \bibinfo{pages}{43--50}.
\newblock


\bibitem[Verbruggen and Snoeck(2023)]%
        {verbruggen2023practitioners}
\bibfield{author}{\bibinfo{person}{Charlotte Verbruggen} {and}
  \bibinfo{person}{Monique Snoeck}.} \bibinfo{year}{2023}\natexlab{}.
\newblock \showarticletitle{Practitioners’ experiences with model-driven
  engineering: a meta-review}.
\newblock \bibinfo{journal}{\emph{Software and Systems Modeling}}
  \bibinfo{volume}{22}, \bibinfo{number}{1} (\bibinfo{year}{2023}),
  \bibinfo{pages}{111--129}.
\newblock


\bibitem[Wan et~al\mbox{.}(2025)]%
        {wan2025structure}
\bibfield{author}{\bibinfo{person}{Bangrui Wan}, \bibinfo{person}{Zheng Wei},
  \bibinfo{person}{Shiyu Wang}, \bibinfo{person}{Jiangping Huang}, {and}
  \bibinfo{person}{Chunqiang Hu}.} \bibinfo{year}{2025}\natexlab{}.
\newblock \showarticletitle{Structure-guided function-level code generation
  with LLMs via UML activity diagrams}.
\newblock \bibinfo{journal}{\emph{Neurocomputing}} (\bibinfo{year}{2025}),
  \bibinfo{pages}{132502}.
\newblock


\bibitem[Wang et~al\mbox{.}(2017)]%
        {wang2017msvl}
\bibfield{author}{\bibinfo{person}{Xiaobing Wang}, \bibinfo{person}{Cong Tian},
  \bibinfo{person}{Zhenhua Duan}, {and} \bibinfo{person}{Liang Zhao}.}
  \bibinfo{year}{2017}\natexlab{}.
\newblock \showarticletitle{MSVL: a typed language for temporal logic
  programming}.
\newblock \bibinfo{journal}{\emph{Frontiers of Computer Science}}
  \bibinfo{volume}{11}, \bibinfo{number}{5} (\bibinfo{year}{2017}),
  \bibinfo{pages}{762--785}.
\newblock


\bibitem[Watanabe et~al\mbox{.}(2025)]%
        {watanabe2025use}
\bibfield{author}{\bibinfo{person}{Miku Watanabe}, \bibinfo{person}{Hao Li},
  \bibinfo{person}{Yutaro Kashiwa}, \bibinfo{person}{Brittany Reid},
  \bibinfo{person}{Hajimu Iida}, {and} \bibinfo{person}{Ahmed~E Hassan}.}
  \bibinfo{year}{2025}\natexlab{}.
\newblock \showarticletitle{On the use of agentic coding: An empirical study of
  pull requests on github}.
\newblock \bibinfo{journal}{\emph{ACM Transactions on Software Engineering and
  Methodology}} (\bibinfo{year}{2025}).
\newblock


\bibitem[Yu et~al\mbox{.}(2025)]%
        {yu2025measuring}
\bibfield{author}{\bibinfo{person}{Liang Yu}, \bibinfo{person}{Emil
  Al{\'e}groth}, \bibinfo{person}{Panagiota Chatzipetrou}, {and}
  \bibinfo{person}{Tony Gorschek}.} \bibinfo{year}{2025}\natexlab{}.
\newblock \showarticletitle{Measuring the quality of generative AI systems:
  Mapping metrics to quality characteristics—Snowballing literature review}.
\newblock \bibinfo{journal}{\emph{Information and Software Technology}}
  \bibinfo{volume}{186} (\bibinfo{year}{2025}), \bibinfo{pages}{107802}.
\newblock


\bibitem[Zhao et~al\mbox{.}(2024)]%
        {zhao2024generating}
\bibfield{author}{\bibinfo{person}{Zelong Zhao}, \bibinfo{person}{Nan Zhang},
  \bibinfo{person}{Bin Yu}, {and} \bibinfo{person}{Zhenhua Duan}.}
  \bibinfo{year}{2024}\natexlab{}.
\newblock \showarticletitle{Generating Java code pairing with ChatGPT}.
\newblock \bibinfo{journal}{\emph{Theoretical Computer Science}}
  \bibinfo{volume}{1021} (\bibinfo{year}{2024}), \bibinfo{pages}{114879}.
\newblock


\end{thebibliography}
\end{document}